\documentclass{article}

\usepackage{arxiv}
\usepackage[utf8]{inputenc} 
\usepackage[T1]{fontenc}    
\usepackage{hyperref}       
\usepackage{url}            
\usepackage{booktabs}       
\usepackage{amsfonts}       
\usepackage{nicefrac}       
\usepackage{microtype}      
\usepackage{lipsum}		
\usepackage{natbib}
\usepackage{doi}
\usepackage{bm}
\usepackage{amsmath}
\usepackage{tcolorbox}
\usepackage{amsbsy}
\usepackage{graphicx}

\title{Towards a Unified Formalism of Multivariate Coefficients of Variation - Application to the Analysis of Polarimetric Speckle Time Series}

\author{  Elise Colin $^1$, Razvigor Ossikovski $^2$ \\
	DTIS-Onera, Palaiseau, France, F-91123, Université de Paris Saclay, \texttt{elise.colin@onera.fr} \\
 LPICM, CNRS, École Polytechnique, Institut Polytechnique de Paris, 91128 Palaiseau, France
}

\date{}
\begin{document}
\maketitle

\begin{abstract}
This article primarily aims to unify the various formalisms of multivariate coefficients of variation, leveraging advanced concepts of generalized means, whether weighted or not, applied to the eigenvalues of covariance matrices. We highlight the existence of an infinite number of these coefficients and demonstrate that they are bounded. Moreover, we link the various coefficients of variation identified in the literature to specific instances within our unified formalism. We illustrate the utility of our method by applying it to a time series of polarimetric radar imagery. In this context, the coefficient of variation emerges as a key tool for detecting changes or identifying permanent scatterers, which are characterized by their remarkable temporal stability. The multidimensionality arises from the diversity of polarizations. The introduction of the various possible coefficients demonstrates how their selection impacts the detection of samples exhibiting specific temporal behaviors and underscores the contribution of polarimetry to dynamic speckle analysis.
\end{abstract}


\section{Introduction}

The coefficient of variation (CV) is a statistical indicator that quantifies the dispersion of data relative to their mean. It is calculated by dividing the standard deviation of the distribution by the mean of that distribution and is often expressed as a percentage. This measure allows for the assessment of the degree of variability in data independently of the data scale, enabling comparisons between data sets with different units of measurement \cite{jalilibal2021monitoring}. The CV is widely used across various research fields to analyze the relative dispersion of data sets, providing insight into the homogeneity or heterogeneity of statistical populations.

The CV proves particularly useful for interpreting the complex and seemingly random patterns generated by speckle. Speckle patterns are typically encountered in coherent imaging systems, such as radar or laser imaging, where interference effects occur due to the interaction of waves with random scatterers or reflective surfaces. 
The coefficient of variation associated with a theoretical distribution of intensities, known as Goodman's speckle, has a fixed value, independent of its mean. In this specific context of use, physicists refer to the coefficient of variation as the speckle contrast. This coefficient of variation can be calculated within a single image by estimating it in a spatial neighborhood of the considered pixel. But it can also be estimated within a time series, which then falls into the application of dynamic speckle. This concept is relevant in both optics and radar, although the use of the term \textit{dynamic speckle} is less common in synthetic aperture radar (SAR).

When calculated temporally within a time series of radar images, the coefficient of variation proves particularly useful for detecting Permanent Scatterers \cite{ferretti2001permanent}. 
Permanent Scatterers are elements within a scene that, due to their temporal stability in terms of backscatter, facilitate the monitoring of displacements or structural deformations with high precision. The coefficient of variation can also be used for identifying significant changes in the observed environment \cite{colin2020change}. In optics, especially in laser imaging techniques, it allows for the analysis of activity or movement within the targeted sample \cite{rabal2018dynamic}. Thus, the coefficient of variation establishes itself as a versatile analytical tool, capable of providing a deep understanding of the phenomena observed across various imaging techniques.

The extension of the concept of the coefficient of variation to the multivariate domain represents a natural and significant evolution of this statistical indicator, tailored to meet the challenges posed by multichannel analyses. The Multivariate Coefficient of Variation (MCV) accounts for the interactions among multiple variables simultaneously, considering the covariance between data channels \cite{jalilibal2021monitoring}. By incorporating inter-variable relationships, the MCV provides another means of analyzing the variability of multidimensional data sets.

In the context of speckle imaging, leveraging the diversity of polarization offers significant benefits. First, in radar, it allows for an increase in the number of detectable Permanent Scatterers \cite{luo2022assessment}. The diversity of polarizations enriches the set of permanent scatterers on which to base deformation estimates \cite{luo2022assessment}. Second, also in radar, polarimetry improves the detection of changes within observed scenes. Changes in the composition, orientation, or state of surfaces result in changes in the behavior of the backscattered wave's polarization. This makes polarimetry particularly suited to detecting changes that might remain undetected in a single polarimetric channel \cite{colin2019}. Finally, integrating the diversity of polarization in laser imaging can increase penetration depth. By selecting specific polarizations, it is possible to minimize reflection effects at the surface of materials, thus allowing light to penetrate more deeply. This is particularly useful for imaging microvascular structures beneath the skin \cite{colin2022imaging}.

Despite significant advancements in the field of speckle imaging and the use of polarimetry, current research reveals a notable gap concerning the plurality of MCV, each with distinct mathematical properties. The examination of current research reveals four main formulations of the MCV, each based on distinct mathematical foundations \cite{aerts2015multivariate,ra7studies,van2005variation,albert2010novel}. These formulations leverage the covariance matrix and the mean vector to calculate variation within a set of vectors. One of the most recent formulation \cite{albert2010novel} has been proposed for its increased robustness to noise and its ability to integrate inter-channel correlation coefficients, reflecting interactions within polarimetric channels.

This diversity of coefficients, rich in analytical potential, raises challenges regarding integration into a coherent analytical framework. Although innovative, the variety of existing approaches often lacks a unified foundation that would allow them to be compared or combined effectively to leverage their specific strengths fully. This situation highlights the need for in-depth reflection and research efforts aimed at consolidating knowledge and analytical tools in this field. It is in this context that our contribution seeks to provide an answer, by proposing a unified formalism that encompasses and connects the different coefficients of variation, thus facilitating their application and interpretation in the multidimensional and polarimetric analysis of speckle images.

Also, in Section 2, we propose to first review the existing coefficients of variation in the literature and their properties.
Then, in Section 3, we show that we can design an infinite number of contrast coefficients, suggesting the potential for better utilization of multichannel information in our measurements. We will focus on illuminating the mathematical properties that interconnect these different coefficients. Through this study, two main categories of coefficients will be examined: those that are equally weighted, referred to as \textit{equally weighted}, and those that are not, termed \textit{non-equally weighted}. We will highlight the merits and limitations of each category and establish the mathematical links that unite them, thus providing a comprehensive understanding and a systematic classification of multivariate contrast measures.

To illustrate the implementation of this formalism and the properties it reveals, we will apply this formalism in Section 4 to radar time series, to examine specific objects such as permanent scatterers and changes. We will discuss the expected results and how they might differ when using the two forms of coefficients. In conclusion, we will summarize the properties identified and the contributions that this work could have on future practices in the field of dynamic speckle, once the measurement exists for multiple polarimetric channels.

\section{State of the Art in Multivariate Coefficients of Variation}

The coefficient of variation is a statistical measure that quantifies the relative variability of a dataset or a distribution. It provides a standardized measure of dispersion, allowing for the comparison of variability between datasets with different means and scales. Several studies have suggested an extension of this definition to include multimodal samples.

Let $\mathbf{P}$ be the matrix of the time-series of $N$ multimodal acquisitions $\mathbf{p}_{k}$ stacked columnwise. Each multimodal acquisition $\mathbf{p}_{k}$ being a \textbf{real} vector of dimension $M$, $\mathbf{P}$ is a real  $M \times N$ matrix: 
\begin{equation}
 \mathbf{P}=\begin{pmatrix}
\mathbf{p}_{1} & ... & \mathbf{p}_{k} & ... & ... &\mathbf{p}_{N} \end{pmatrix}= \begin{pmatrix}
{p_1}_{1} & ... & {p_1}_{k} & ... & ... &{p_1}_{N}  \\ {p_2}_{1} & ... & {p_2}_{k} & ... & ... & {p_2}_{N} \\ ... & ... & ... & ... &... & ... \\{p_{M}}_{1} & ... & {p_{M}}_{k} & ... & ... &{p_{M}}_{N} \end{pmatrix}
\end{equation}

We can then define the mean vector $\bm{\mu}$ of dimension $M$, 

\begin{equation} 
 \bm{\mu}=\begin{pmatrix} \mu_1 \\ \mu_2\\... \\ \mu_{M} \end{pmatrix} = \frac{1}{N} \displaystyle \sum_{k=1}^N \mathbf{p}_{k} = \begin{pmatrix} \frac{1}{N} \displaystyle \sum_{k=1}^N {p_1}_k \\ \frac{1}{N} \displaystyle \sum_{k=1}^N {p_2}_k\\... \\ \frac{1}{N} \displaystyle \sum_{k=1}^N {p_{M}}_k \end{pmatrix} = \mathbf{P}\begin{pmatrix} \frac{1}{N} \\ \frac{1}{N}\\... \\ \frac{1}{N} \end{pmatrix}
\end{equation}

as well as the $M \times M$ covariance matrix $\mathbf{C}$,

\begin{equation} 
\mathbf{C}=\frac{1}{N}\displaystyle \sum_{k=1}^N (\mathbf{p}_{k}- \bm{\mu})(\mathbf{p}_{k}- \bm{\mu})^t=\frac{1}{N}\mathbf{P}\mathbf{P}^t-\bm{\mu}\bm{\mu}^t
\end{equation}

Based on the above definitions, there exist at least four mathematical formulations of MCV in the literature,

\begin{equation}
  \gamma_{\textrm{R}}=\sqrt{\frac{\det(\mathbf{C})^{\frac{1}{M}}}{\mathbf{\bm{\mu}^{t} \bm{\mu}}}}, ~~\gamma_{\textrm{VV}}=\sqrt{\frac{\textrm{trace}(\mathbf{C})}{\mathbf{\bm{\mu}^{t} \bm{\mu}}}},  ~~   \gamma_{\textrm{VN}}=\sqrt{\frac{1}{\mathbf{\bm{\mu}^{t} C^{-1}\bm{\mu}}}}, ~~\gamma_{\textrm{AZ}}=\sqrt{\frac{\mathbf{\bm{\mu}^{t} C \bm{\mu}}}{(\mathbf{\bm{\mu}^{t} \bm{\mu})^2}}}
\end{equation}

Note that all these formulations are invariant with respect to orthogonal transformations of basis. They are commonly defined for real acquisitions however, they remain applicable for complex acquisitions as well provided the transpose is replaced by the complex conjugate transpose; this is so owing to the semi-positive definiteness of the covariance matrix. The only condition for their validity is that the mean vector $\bm{\mu}$ be non-zero.

The coefficient $\gamma_{\textrm{R}}$ was proposed by \cite{ra7studies} in the 1960s. It suffers from the drawback of rapidly approaching zero when any of the eigenvalues of the matrix $\textbf{C}$ is close to zero. The definition proposed by \cite{van2005variation} for $\gamma_{\textrm{VV}}$ does not have this drawback but it does not take into correlations between dimensions. The coefficient $\gamma_{\textrm{VN}}$, proposed in \cite{voinov2012unbiased}, accounts for possible correlations, but its value is often noisy on real data as its definition involves the inverse of $\textbf{C}$.
Finally, the paper \cite{albert2010novel} advanced the last-to-date formulation of the MCV $\gamma_{\textrm{AZ}}$. The novel coefficient definition aims at removing certain limitations of the previous definitions, as well as at enhancing the MCV applicability to various fields of study. The novel formulation takes into account the correlations and interdependencies among the variables, thus providing a more comprehensive measure of their variability. The paper also provides mathematical derivations and analytical insights to support the proposed formulation. Moreover, it reports practical applications and examples demonstrating the usefulness and effectiveness of the novel MCV in real-world scenarios.

\cite{aerts2015multivariate,albert2010novel} studied the different definitions and properties of the MCV, including their robustness to outliers, efficiency in estimating variability, and sensitivity to changes in the data distribution. They evaluated and compared these measures using both simulated data and real-world datasets, providing empirical evidence for their performance. \cite{aerts2017robust} emphasizes the importance of choosing the most suitable measure based on the specific requirements and characteristics of the data at hand, by taking into account factors such as data structure, dimension, and robustness considerations. 

In more recent studies \cite{ditzhaus2023inference,ditzhaus2022permutation}, the authors investigated not only the effect of each piece of information individually but also the effect of their interaction through a factorial design. The inference methods for the MCV in this context help to statistically test whether there is a significant difference in the variability of the data across groups or experimental conditions, and provide a basis for further analyses after a general difference has been detected, to specifically explore where these differences lie (for example, between which groups or under which conditions the variability differences are most pronounced). However, the paper deals with the classical MCV definitions from the literature and does not propose any novel ones.

In the following section, we advance two generic formulations of the MCV that encompass the various definitions proposed so far.

\section{Unified Definitions of the Multivariate Coefficients of Variation and Their Properties}

\subsection{Contrast definition based on the power mean}

By definition, the covariance matrix  $\mathbf{C}$ is real symmetric (or Hermitian, in the complex case) and semi-positive definite, i.e. it has only positive or zero eigenvalues.

We can thus evaluate the power mean (or generalized mean) of the non-negative eigenvalues of the covariance matrix. If $q$ is a non-zero real number and {$\displaystyle \lambda_{1},\dots ,\lambda_{M}$} are the eigenvalues of $\mathbf{C}$ then: 

\begin{equation}
    m_q(\lambda_{1},\dots ,\lambda_{M})=\left(\frac{1}{M}\sum_{i=1}^M \lambda_i^q \right)^{\frac{1}{q}} = \left(\frac{1}{M} \textrm{trace}(\mathbf{C}^q) \right)^{\frac{1}{q}}.
\end{equation}

If $q = 0$ then the approach to the limit yields:

\begin{equation}
    m_0(\lambda_{1},\dots ,\lambda_{M})=\left(\prod_{i=1}^M \lambda_i \right)^{\frac{1}{M}} = \left( \textrm{det}(\mathbf{C}) \right)^{\frac{1}{M}}.
\end{equation}

A few values of $q$ yield familiar special cases: $q = 0$ above corresponds to the expression of the geometric mean, $q = -1$: to the harmonic mean, $q = 1$ to the arithmetic mean, and $q = 2$ to the quadratic mean (or root mean square).

The power mean has properties that will be useful later namely, 
the generalized mean inequality states that if $q_1<q_2$, then $ m_{q_1}\leq m_{q_2}$, and also that $\min{\lambda_i}\leq\ m_q\leq\max{\lambda_i}$ for any real $q$.

We are now in a position to propose the following generic form of the MCV, which we will call Equally-Weighted Contrast (EWC):

\begin{equation}
\gamma_{q}^{eq} = \frac{\sqrt{m_q}}{\mu} 
\end{equation}

where $\mu=\sqrt{{\bm{\mu}^{t} \bm{\mu}}}$. This definition is valid in the complex case too, provided the complex conjugate transpose instead of the (real) transpose is used. For $q = 0$ and $q = 1$ we get respectively:

$$\gamma_{0}^{eq}=\gamma_{\textrm{R}},~~ \gamma_{1}^{eq} = \frac{\gamma_{\textrm{VV}}}{\sqrt{M}},$$

i.e. we recover the first two MCV definitions as special cases. Furthermore, in virtue of the generalized mean inequality we have:

$$\gamma_{\textrm{R}} \leq \frac{\gamma_{\textrm{VV}}}{\sqrt{M}} \leq \gamma_{\textrm{VV}}, $$

in accordance with the inequality $\gamma_{\textrm{R}} \leq \gamma_{\textrm{VV}}$ established in \cite{albert2010novel} in a different way.

In principle, the generalized contrast definition is valid for any real $q$ (not only for an integer one), potentially generating an infinite number of contrasts. If $q$ takes a series of integer values, then we get the following series of inequalities from the generalized mean inequality,  

$$\gamma_{-\infty}= \frac{\sqrt{\lambda_{\textrm{min}}}}{\mu} \leq ... \leq \gamma_{-1}^{eq} \leq \gamma_{0}^{eq} \leq \gamma_{1}^{eq} \leq ...\leq \frac{\sqrt{\lambda_{\textrm{max}}}}{\mu}=\gamma_{\infty}$$

The equalities are reached when all eigenvalues are equal.

\subsection{Contrast definition based on the weighted power mean}

We now turn to the evaluation of the weighted power mean of the eigenvalues of the covariance matrix by using the squared components of the mean vector, expressed in the basis into which the covariance matrix is diagonal, as weights.

Specifically, let $ \mathbf{D}=\mathbf{U}^{-1}\mathbf{C}\mathbf{U}=\mathbf{U}^{t}\mathbf{C}\mathbf{U} $ be the transformation by the orthogonal matrix $\mathbf{U}$ diagonalizing $\mathbf{C}$ to $ \mathbf{D}=\textrm{diag}(\lambda_1, \dots, \lambda_{M})$. Then $\bm{\mu'}=\mathbf{U}^{-1}\bm{\mu}=\mathbf{U}^{t}\bm{\mu}$ is the mean vector in the basis into which $\mathbf{C}$ is diagonal. Furthermore, we have $\bm{\mu'}^{t}\mathbf{D}^q \bm{\mu'}=\bm{\mu}^{t}\mathbf{C}^q \bm{\mu} $ and $ \bm{\mu'}^{t} \bm{\mu'}=\bm{\mu^{t}} \bm{\mu} =\mu^2$ by the property of orthogonal transformations.

Now, if $q$ is a non-zero real number, and {$\displaystyle \lambda_{1},\dots,\lambda_{M}$} are the eigenvalues of $\mathbf{C}$ then the weighted power mean is:  

\begin{equation}
    m_q^{w}({\mu'}_1,\dots, {\mu'}_{M},\lambda_{1},\dots ,\lambda_{M})=\left( \frac{\displaystyle\sum_{i=1}^M {\mu}_i^{'2}\lambda_i^q  }{\displaystyle\sum_{i=1}^M {\mu}_i^{'2}}  \right)^{\frac{1}{q}}=\left( \frac{\bm{\mu'^{t}}\mathbf{D}^q \bm{\mu'}}{\bm{\mu'^{t}\mu'}}  \right)^{\frac{1}{q}}=\left( \frac{\bm{\mu^{t}}\mathbf{C}^q \bm{\mu}}{\bm{\mu^{t}\mu}}  \right)^{\frac{1}{q}} 
\end{equation}

If $q=0$ the approach to the limit provides: 
\begin{equation}
    m_0^{w}({\mu'}_1,\dots, {\mu'}_{M},\lambda_{1},\dots ,\lambda_{M})=\left(\prod_{i=1}^M \lambda_i^{{\mu'_i}^2} \right)^{\displaystyle\frac{1}{\displaystyle\sum_{i=1}^M {\mu}_i^{'2}}}= \exp{\left(\frac{\bm{\mu'^{t}}\ln{\mathbf{D}} \bm{\mu'}}{\bm{\mu'^{t}\mu'}}\right)}= \exp{\left(\frac{\bm{\mu^{t}}\ln{\mathbf{C}} \bm{\mu}}{\bm{\mu^{t}\mu}}\right)}
 \end{equation}

where $\ln$ stands for the matrix logarithm, i.e. $ \ln{\mathbf{C}}=\mathbf{U}\ln{\mathbf{D}}\mathbf{U}^{-1}=\mathbf{U}\textrm{diag}(\ln{\lambda_1}, \dots, \ln{\lambda_{M}})\mathbf{U}^{-1} $ . The weights in the above means are thus the squared components of the $\bm{\mu'}$ vector. Both expressions hold also in the complex case, by squaring the absolute values of the components of $\bm{\mu'}$ to be used as weights, and by taking complex conjugate transposition instead of real one. The transformation matrix $\mathbf{U}$ will then be unitary instead of orthogonal.

We can now define the Non-Equally-Weighted Contrast (NEWC) in full analogy to its equally weighted counterpart as 

\begin{equation}
    \gamma^{ne}_q =  \frac{\sqrt{m_q^{w}}}{\mu} 
\end{equation}

We can then recover the remaining two of the conventional MCV as special cases,

  $$\gamma_{1}^{ne} =  \gamma_{\textrm{AZ}}  \hspace{2cm} \gamma_{-1}^{ne} =  \gamma_{\textrm{VN}}$$

As with its equally weighted counterpart, the generalized non-equally weighted contrast definition is valid for any real $q$ (not only for an integer one), potentially generating an infinite number of contrasts. For a series of integer values of $q$ we get, as before, the series of inequalities,

$$\gamma_{-\infty}=\frac{\sqrt{\lambda_{\textrm{min}}}}{\mu} \leq ... \leq \gamma_{-1}^{ne} \leq \gamma_{0}^{ne} \leq \gamma_{1}^{ne} \leq ...\leq \frac{\sqrt{\lambda_{\textrm{max}}}}{\mu}=\gamma_{\infty}$$

In particular, we see that $ \gamma_{\textrm{VN}} \leq \gamma_{\textrm{AZ}} $.

Finally, notice that both the equally and non-equally weighted contrasts can be further generalized if the generalized $f$-mean instead of the power mean is used namely,

\begin{equation}
    m^f(\lambda_{1},\dots ,\lambda_{M})=f^{-1}\left(\frac{1}{M}\sum_{i=1}^M f(\lambda_i) \right) = f^{-1}\left(\frac{1}{M} \textrm{trace}(f(\mathbf{C})) \right)
\end{equation}

and

\begin{equation}
    m^{fw}({\mu'}_1,\dots, {\mu'}_{M},\lambda_{1},\dots ,\lambda_{M})=f^{-1}\left( \frac{\displaystyle\sum_{i=1}^M {\mu}_i^{'2}f(\lambda)  }{\displaystyle\sum_{i=1}^M {\mu}_i^{'2}}  \right)=f^{-1}\left( \frac{\bm{\mu'^{t}}f(\mathbf{D}) \bm{\mu'}}{\bm{\mu'^{t}\mu'}}  \right)=f^{-1}\left( \frac{\bm{\mu^{t}}f(\mathbf{C}) \bm{\mu}}{\bm{\mu^{t}\mu}}  \right) 
\end{equation}

where $f$ is a real continuous monotonic function and $f^{-1}$ is its inverse function. If $f$ is also a monotonic function of some real parameter $q$ then a series of inequalities between contrasts, analogous to those already reported, will hold. The power mean is obviously obtained for $f=x^q$; in particular, taking the limit for $q = 0$ is equivalent to setting $f = \textrm{ln}x$. Eventually, the terms "equally weighted" and "non-equally weighted" also become clear: the former is obtained from the latter when all squared components of the mean vector $\bm{\mu'}$ used as weights are equal.

\section{Results on radar time-series and discussion}

\subsection{Motivation}

In existing literature, tests on the coefficients of variation typically rely on statistical data. However, this paper seeks to explore their application in a more practical context, specifically in the analysis of radar speckle images. The coefficient of variation in radar time series serves dual purposes: firstly, for identifying permanent scatterers having low CV, as outlined in \cite{colin2021urban}, and secondly, for change detection according to high CV, as discussed in \cite{colin2020change}. Given the dynamic nature of urban areas—marked by continuous development, construction, and land cover modifications—the study validates the CV as an effective metric for identifying changes through the SAR time series. The method's reliability is confirmed by comparing its outcomes with ground truth data, such as land cover maps or reference images, ensuring its accuracy and practical utility.

Radar polarimetry is a sophisticated technique used in radar systems that involves the measurement and analysis of the polarization state of electromagnetic waves. This method extends traditional radar by not only measuring the intensity of the backscattered signal but also examining how the wave's electric field vector oscillates over time. Polarization, in this context, refers to the orientation of the electromagnetic wave's electric field vector in the plane perpendicular to the direction of wave propagation. In radar polarimetry, the transmitted and received signals can be polarized in various ways, typically described as horizontal (H) or vertical (V) polarization, although circular and elliptical polarizations can also be used. By emitting and receiving signals with different polarizations, polarimetric radar systems can obtain more information about the target than can be gleaned from the signal's intensity alone. This additional information includes insights into the target's geometry, orientation, and material properties, as well as environmental factors such as moisture content.

\cite{colin2019} demonstrate that diversity in polarizations significantly enhances detection accuracy. This is because deterministic objects usually exhibit a preferred orientation, and consequently, a specific polarimetric response. The author highlighted that across continental regions, a full polarimetric acquisition was essential for maximizing change detection, noting that omitting a polarimetric channel results in losing at least 25\% of detections. Thus, to adapt change detection algorithms based on the coefficient of variation for polarimetric data, they proposed to use the highest coefficient of variation across all available polarimetric channels. 

Our contribution extends this approach by analyzing various MCV. To facilitate a comparative analysis of two distinct approaches to image acquisition, we examine scenarios within the realm of polarimetric radar measurements:

In the first scenario, known as \textit{\textbf{full polarimetry}}, the complete polarimetric scattering matrix is captured. This involves the transmitter emitting signals successively in both Horizontal (H) and Vertical (V) polarizations, with the receiver correspondingly collecting data in these two polarizations. This comprehensive method ensures that all possible polarization interactions are recorded, providing a detailed characterization of the target's scattering properties.

The second scenario, referred to as \textbf{\textit{partial polarimetry}}, involves acquiring only a subset of the polarimetric scattering matrix. An example of this could be the collection of data only for the polarization pairs HH (Horizontal transmit, Horizontal receive) and HV (Horizontal transmit, Vertical receive). This approach, while not as exhaustive as full polarimetry, still offers valuable insights into the scattering behavior of targets but with reduced data complexity and acquisition time.

\subsection{First Case Study: A Fully Polarimetric UAVSAR Time Stack}

For each pixel within this region, we construct a vector \textbf{p}, characterized by three components corresponding to the amplitude values recorded in each of the polarimetric channels: HH (Horizontal transmit, Horizontal receive), HV (Horizontal transmit, Vertical receive), and VV (Vertical transmit, Vertical receive). Note here that we consider the magnitudes of the electric fields because, although the formalism of MCVs can be extended to complex cases, in practice and theoretically, complex field vectors follow circular Gaussian statistics \cite{goodman2015statistical}, and therefore have an average magnitude equal to zero.

From these $N$ temporal observations of dimension $M=3$ at each pixel, we estimate the coherence matrix resulting in a 3x3 matrix. Likewise, the mean vector, denoted as $\bm{\mu}$, is established to have a three-dimensional structure, reflecting the average scattering response in the aforementioned polarimetric channels.

\begin{figure}
    \centering
    \includegraphics[height=6cm]{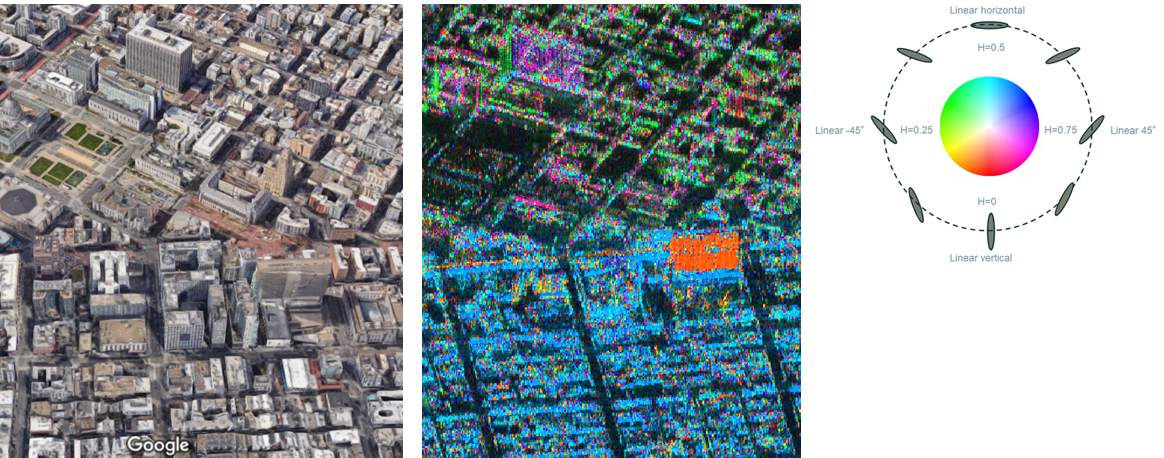}
    \caption{Left: Optical View of the neighborhood selected over San Francisco- Right: polarimetric representation of radar image selected. Hue represents the polarimetric orientation angle, Saturation is the temporal degree of polarization, and amplitude is the time-filtered amplitude.}
        \label{fig:Sanfrancisco}
\end{figure}

An optical view of the Selected Neighborhood covered by the image is shown in Figure \ref{fig:Sanfrancisco} on the left. This image provides a visual representation of the area under study, offering a conventional optical perspective to facilitate geographical and contextual orientation. The view is located at the interface of the SOMA district—a neighborhood in San Francisco, California, just south of Market Street. In the right panel, we give a polarimetric representation of the Selected area. This image is rendered using polarimetric radar data, where the hue corresponds to the polarimetric orientation angle, indicating the dominant scattering mechanism direction. The saturation level reflects the degree of polarization. The brightness level represents the amplitude, which has been filtered over time to enhance the signal-to-noise ratio. 

Notably, two buildings within this visualization emerge with distinctive polarimetric signatures due to their azimuth orientations: the San Francisco Federal Building, prominently displayed in red towards the center-right of the image, and the San Francisco Passport Agency, positioned at the top of the image. The latter features the projections of two facades, each highlighted in distinct colors of green and magenta, respectively, showcasing the diverse polarimetric responses influenced by architectural design and orientation.

\begin{figure}
    \centering
    \includegraphics[width=12cm]{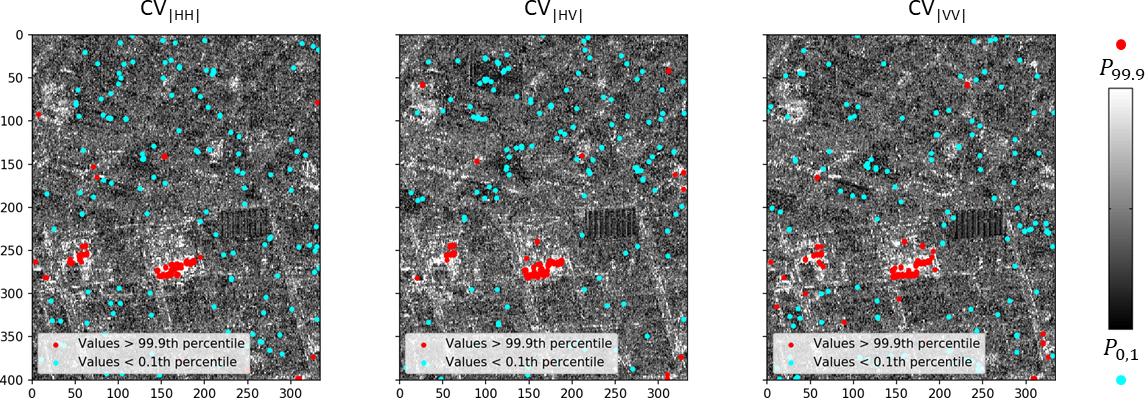}
        \caption{Representation of the three CV computed for each polarimetric channel: CV$_{|\textrm{HH}|}$, CV$_{|\textrm{HV}|}$ and CV$_{|\textrm{VV}|}$, and Overlay of detected pixels having respectively the 1/1000 lowest values (in blue) and the 1/1000 highest values (in red) over the (|HH|,|HV|,|VV|) time-series over San Francisco.}
     \label{fig:detection_san_francisco_lexico}
\end{figure}

Figure \ref{fig:detection_san_francisco_lexico} displays the variation coefficient maps calculated in the polarimetric channels HH, HV, and VV, highlighting the top 1/1000 points in red and the lowest 1/10000 points in blue. Consistent with the rest of this article, the display of the maps is adjusted between the values of the first and 99th percentiles to ensure qualitatively accurate visualization of dynamics. Overall, the red points are primarily located on two buildings undergoing construction along Market Street. We can verify, through archival optical images, that these two areas are indeed under construction at the dates of acquisition in 2018.
Permanent Scatterers, however, are scattered throughout the image, with some minor differences observed among the three images.

\begin{figure}
    \centering
    \includegraphics[width=12cm]{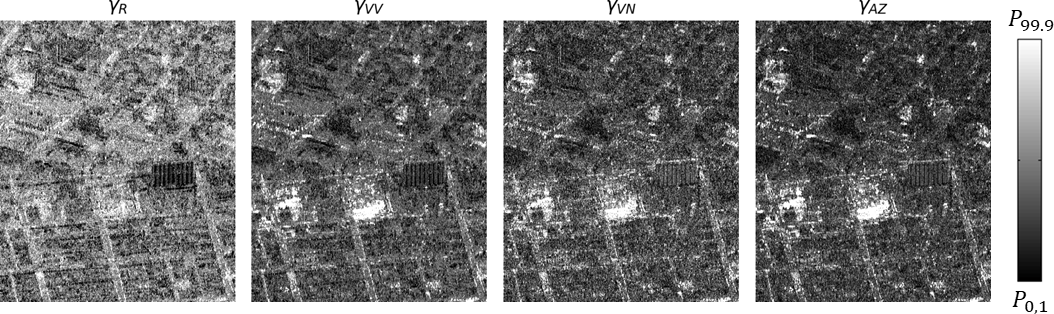}
        \caption{Representation of the 4 classical MCV of the literature, over the (|HH|,|HV|,|VV|) time series over San Francisco.
    $\gamma_{\textrm{R}}=\gamma_{0}^{eq}$, $\gamma_{\textrm{VV}}=\gamma_{1}^{eq}$, $\gamma_{\textrm{VN}}=\gamma_{-1}^{ne}$, $\gamma_{\textrm{AZ}}=\gamma_{1}^{ne}$}
        \label{fig:cv_san_francisco_classical}
\end{figure}

\begin{figure}
    \centering
    \includegraphics[width=10cm]{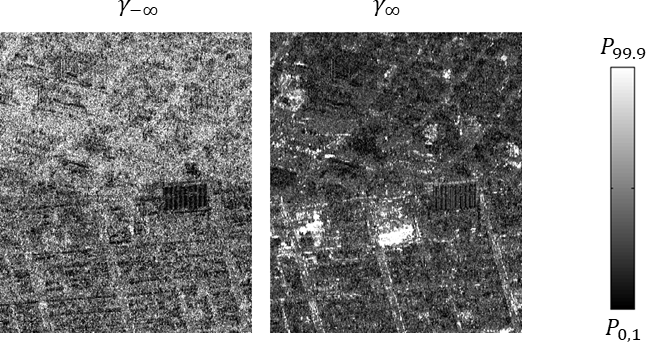}
    \caption{The two images correspond to the lower and upper limit values of the proposed MCVs, derived from the minimum and maximum eigenvalues of the covariance matrix, over the (HH,HV,VV) time series over San Francisco.}
    \label{fig:lminmax_san_francisco}
\end{figure}

Figure \ref{fig:cv_san_francisco_classical} presents the four classic values of multivariate variation coefficients  $\gamma_{\textrm{R}}=\gamma_{0}^{eq}$, $\gamma_{\textrm{VV}}=\gamma_{1}^{eq}$, $\gamma_{\textrm{VN}}=\gamma_{-1}^{ne}$, and $\gamma_{\textrm{AZ}}=\gamma_{1}^{ne}$. The one corresponding to $\gamma_{\textrm{R}}$ is qualitatively closer to that of the lower limit convergence. The other three are more aligned with the upper convergence. Figure \ref{fig:lminmax_san_francisco} shows the map of limit variation coefficients, calculated from the minimum and maximum wavelengths of the covariance matrix. This time, the difference is stark. The image of maximum convergence reveals changes that the minimal value image struggles to detect. Conversely, the latter better highlights permanent-type targets.

\begin{figure}
    \centering
    \includegraphics[width=12cm]{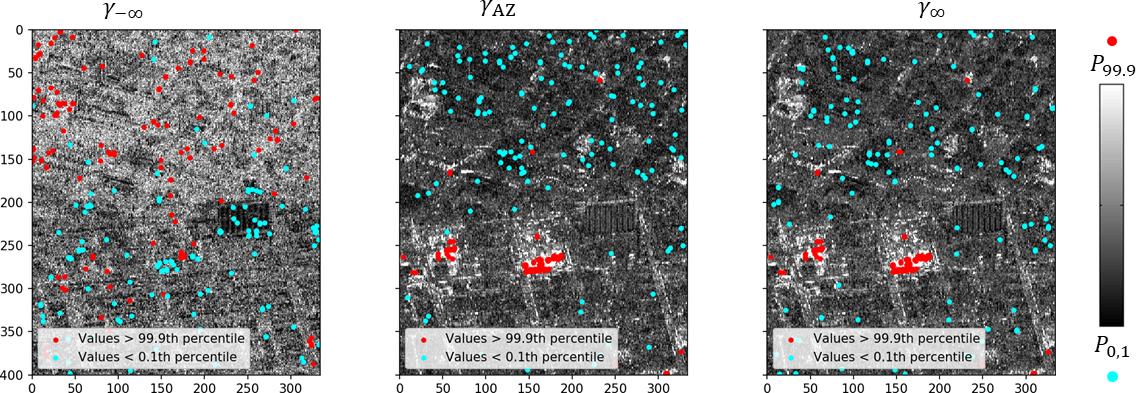}
        \caption{Representation of three possible MCV: $\gamma_{-\infty}$, $\gamma_{\textrm{AZ}}=\gamma_{1}^{ne}$ and $\gamma_{-\infty}$, and Overlay of detected pixels having respectively the 1/1000 lowest values (in blue) and the 1/1000 highest values (in red) over the (|HH|,|HV|,|VV|) time-series over San Francisco.}
    \label{fig:detection_san_francisco}
\end{figure}

Finally, Figure \ref{fig:detection_san_francisco} demonstrates that the choice of MCV significantly influences the detection that can be inferred. In this figure, we can identify points that are detected among the lowest values as permanent scatterers in the image of the lower CV $\gamma_{-\infty}$, while these same points are detected among the highest values as changes according to $\gamma_{\textrm{AZ}}=\gamma_{1}^{ne}$ and $\gamma_{-\infty}$. By analyzing these particular pixels, we can see that they correspond to coherence matrices of rank close to zero. This effect had already been noted for $\gamma_{\textrm{R}}$, which rapidly drops to values close to zero as soon as one of the eigenvalues is also close to zero. This means that for temporal profiles highlighting a change not evident in all polarimetric channels, MCVs of sufficiently low order will be able to detect them as permanent scatterers, by focusing on the polarimetric channels that do not perceive the change. Conversely, the changes detected by these same MCVs are evident in all polarimetric channels.

\subsection{Second Illustration Case: A partial Polarimetric Tandem-X time-series}

\begin{figure}
    \centering
    \includegraphics[height=5cm]{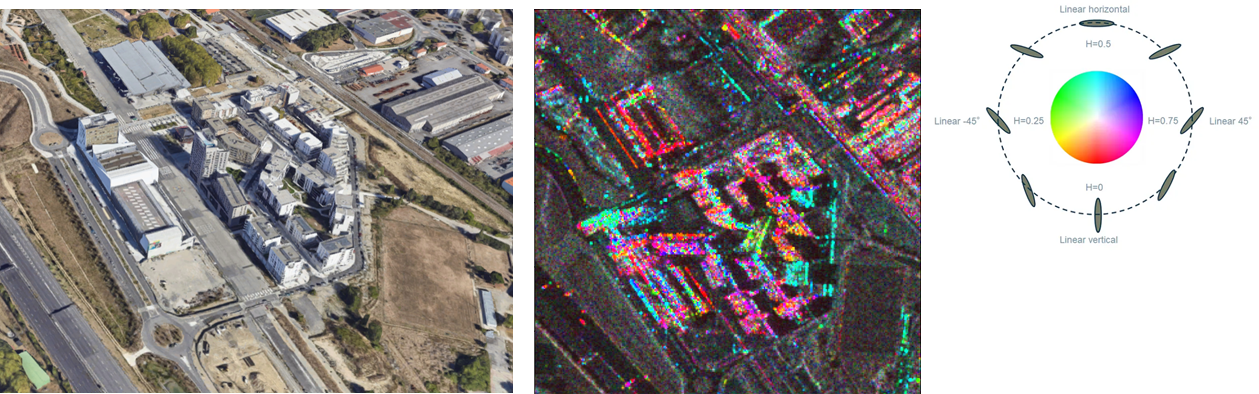}
    \caption{Left: Optical View of the neighborhood selected over Toulouse- Right: polarimetric representation of radar image selected. Hue represents the polarimetric orientation angle, Saturation is the temporal degree of polarization, and amplitude is the time-filtered amplitude.}
    \label{fig:Toulouse}
\end{figure}

In our second scenario, we considered a stack of TerraSAR-X images, kindly acquired by the DLR over Toulouse (France), in HH and HV polarizations. Here, we selected 10 dates between May 2021 and October 2022. Once again, we focused on a small sub-area of 300x300 pixels containing numerous changes, around the Halle de la Machine. The corresponding optical and polarimetric radar images are shown in Figure \ref{fig:Toulouse}.

\begin{figure}
    \centering
    \includegraphics[width=12cm]{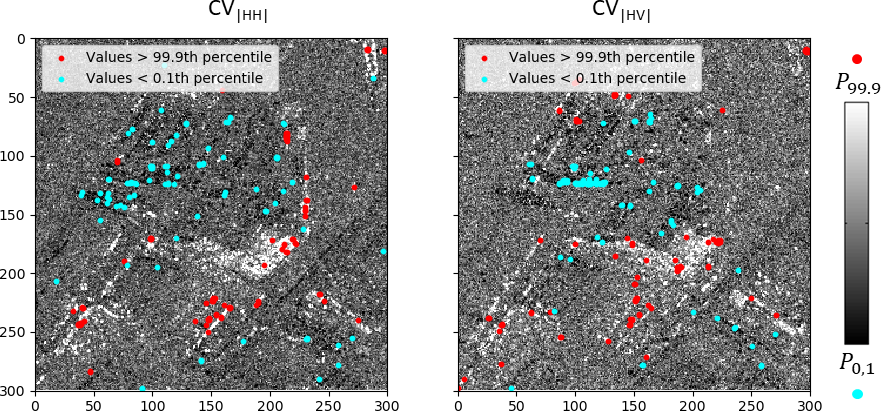}
            \caption{Representation of the two CV computed for each polarimetric channel: $\gamma_{\textrm{HH}}$ and $\gamma_{\textrm{HV}}$, and Overlay of detected pixels having respectively the 1/1000 lowest values (in blue) and the 1/1000 highest values (in red) over the (|HH|,|HV|) time-series over Toulouse.}
                \label{fig:detection_toulouse_classic}
\end{figure}

In this scenario, we have images in two polarization channels, HH and HV. The variation coefficients shown in Figure \ref{fig:detection_toulouse_classic} reveal several differences. The HH polarization can detect changes where points are aligned along the azimuth axis of the image, whereas these same points are not detected in HV polarization. The opposite trend is observed for the permanent scatterers: some detected points are aligned along the range axis in HV, they do not appear in HH. Overall, there is no correspondence of detected points within the two criteria calculated for each of the polarization channels.

\begin{figure}
    \centering
    \includegraphics[width=12cm]{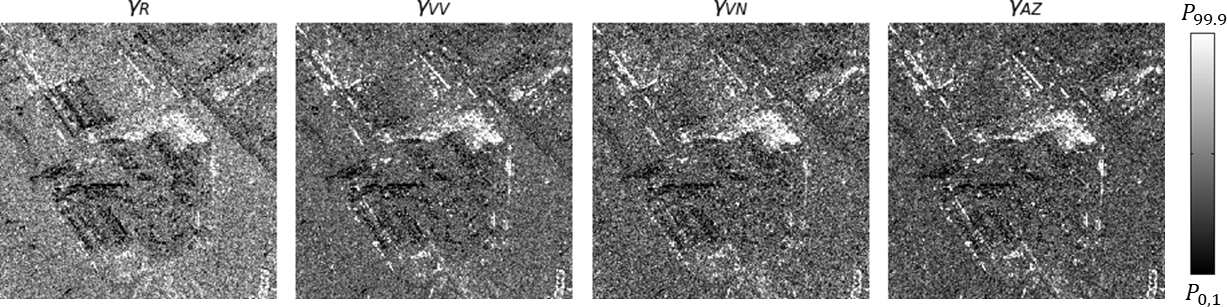}
    \caption{Representation of the 4 classical MCV of the  litterature, over the (HH,HV) time series over Toulouse.
    $\gamma_{\textrm{R}}=\gamma_{0}^{eq}$, $\gamma_{\textrm{VV}}=\gamma_{1}^{eq}$, $\gamma_{\textrm{VN}}=\gamma_{-1}^{ne}$, $\gamma_{\textrm{AZ}}=\gamma_{1}^{ne}$}
 \label{fig:cv_toulouse_classical}
\end{figure}

The four classical variation coefficients shown in Figure \ref{fig:cv_toulouse_classical} behave similarly to those discussed in the previous section on San Francisco.

\begin{figure}
    \centering
    \includegraphics[width=10cm]{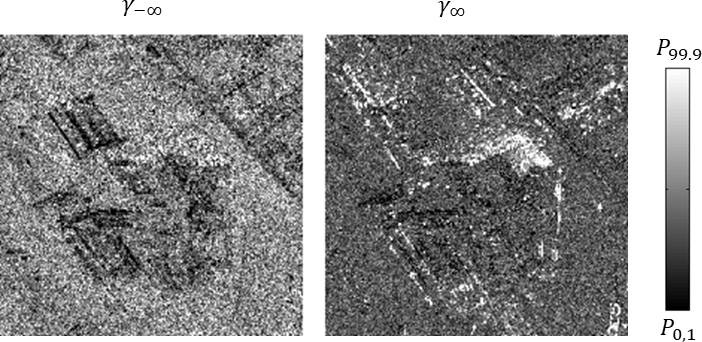}
 \caption{The two images correspond to the lower and upper limit values of the proposed MCVs, derived from the minimum and maximum eigenvalues of the covariance matrix, over the (HH,HV) time series over Toulouse.}
     \label{fig:lminmax_toulouse}
\end{figure}

\begin{figure}
    \centering
    \includegraphics[width=12cm]{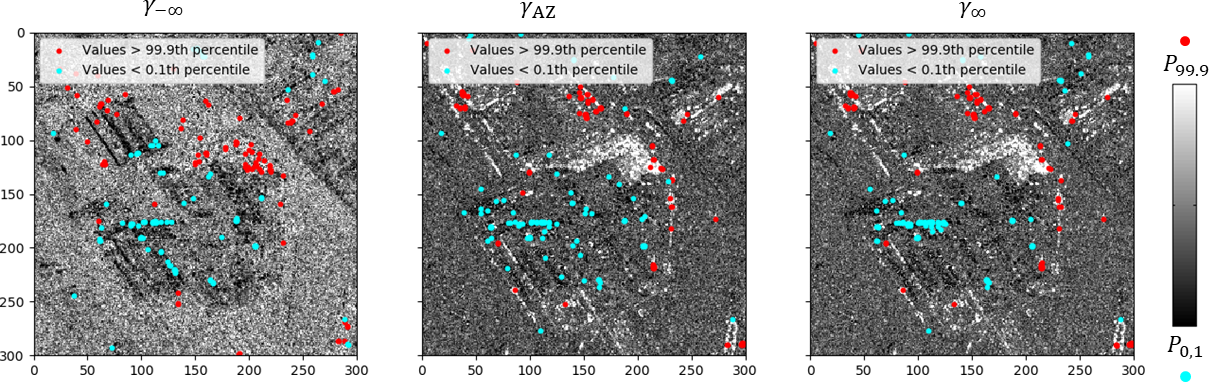}
            \caption{Representation of three possible MCV: $\gamma_{-\infty}$, $\gamma_{\textrm{AZ}}=\gamma_{1}^{ne}$ and $\gamma_{-\infty}$, and Overlay of detected pixels having respectively the 1/1000 lowest values (in blue) and the 1/1000 highest values (in red) over the (HH,HV) time-series over Toulouse.}
                \label{fig:detection_toulouse}
\end{figure}

Figure \ref{fig:lminmax_toulouse} confirms these findings through the two extreme cases, as does the detection figure \ref{fig:detection_toulouse}. The $\gamma_{\textrm{R}}$ coefficient highlights more structuring in the low values. Qualitatively, the 3 other coefficients look similar but reveal some differences when observed more closely.
The two limit coefficients represented in Figure \ref{fig:lminmax_toulouse} reveal notable differences. Many changes are barely perceptible in the lower limit. Yet, the detection in Figure \ref{fig:detection_toulouse} from the lower limit MCV still captures the main detection, which the other coefficients ignore, by focusing more on other areas of the image.

The detection in figure \ref{fig:detection_toulouse} reveals that qualitative analysis is insufficient to perceive on which pixels the coefficient will take its extreme values. A priori, one might think that the MCVs associated with low q values will better reveal changes that are illustrated in all polarimetric channels, but on the other hand, will capture permanent scatterers that correspond to stable signals in at least one of the channels. In contrast, MCVs associated with high q values will better detect changes that are illustrated in only one channel, and permanent scatterers that will be stable across all channels.

\section{Conclusion}

In this paper, we have demonstrated that it is possible to propose an infinity of formulations to define a multivariate coefficient of variation. The theoretical framework proposed for their definition is based on the concept of generalized means applied to the eigenvalues of the covariance matrix, weighted or not by the values of the mean vector calculated in the diagonalization basis. We have shown that the four coefficients traditionally used in the literature are particular cases of this unified formalism. We have revealed the inequalities that apply to them and expressed the lower and upper limit values of these coefficients.

We applied this formalism to the calculation of different MCVs on two polarimetric time series of SAR images. These illustrations show that the choice of the MCV used as a criterion for detecting stable features or changes affects the detection outcome. It can be surmised that MCVs of lower ranks are more adept at detecting changes visible in all available polarizations, while MCVs of higher ranks are better suited to detect changes visible in only one polarimetric channel, or permanent scatterers stable across all polarimetric channels. Further analysis will refine the understanding of the different practical limit cases encountered, to better guide the choice of MCV.

\subsection*{Acknowledgements}

First and foremost, we would like to express our sincere gratitude to Pr. Adelin Albert for graciously sharing his work with me. We would also like to thank Dr. Aurélien Plyer (Onera), who helped in developing Python tools capable of processing our dynamic speckle data, and to Dr. Enrique Garcia-Caurel (LPICM,CNRS) whom we thank for all the fruitful scientific discussions around this topic. These efforts are part of the MUSIC Chair project, which is funded by ONERA. 

The UAVSAR data used in this study were kindly provided by the NASA Jet Propulsion Laboratory. We would like to express our gratitude for access to these valuable data, which were essential for the completion of this work. The TanDEM-X data were provided by DLR under the scientific proposal OTHER0103. This work would not have
been possible without the collaboration of Paola Rizzoli and Jose Luis Bueso Bello in DLR, whom we thank very
sincerely. This fruitful collaboration was initiated in the framework of the ONERA-DLR virtual research center for
cooperation in AI and applications in aerospace engineering.

\subsection*{Contributions}
E.C initiated the investigation of multivariate coefficients of variation for the analysis of polarimetric speckle time series, while R.O. pioneered the concept of unifying these multivariate coefficients of variation. Both have contributed to the writing of the article.

\bibliographystyle{unsrtnat}
\bibliography{sample} 

\end{document}